\documentclass[12pt]{article}

\usepackage{setspace}
\usepackage{authblk}
\usepackage{amssymb}
\usepackage{amsfonts, amssymb,amsthm,amsmath, graphics,color,lscape,bm,float,graphicx, subfig}
\usepackage{multirow}

\usepackage[]{algorithm2e}
\usepackage{titling} % Customizing the title section
\usepackage{mathtools}

\numberwithin{equation}{section}
\theoremstyle{plain}

\usepackage{tikz}
\usetikzlibrary{arrows,decorations.pathmorphing,backgrounds,positioning,fit,petri}
\usetikzlibrary{shapes}
\usepackage{dirtytalk}

\title{\textbf{ Causal Impact of Web Browsing and Other  Factors on Research Publications}}

\author{ \textbf{Bharathi Manjula .K}$^{a}$, \textbf{Sourish Das}$^{b}$, and \textbf{Jehadeesan .R}$^{c}$}

\date{$^{a}$ Homi Bhabha National Institute\\
$^{b}$ Chennai Mathematical Institute\\
$^{c}$ Indira Gandhi Centre for Atomic Research\\
\vspace{0.25cm}
\today}

\begin{document}

\maketitle

\begin{abstract}
         In this paper, we study the causal impact of the web-search activity on the research publication. We considered observational prospective study design, where research activity of 267  scientists is being studied. We considered the Poisson and negative binomial regression model for our analysis. Based on the Akaike's Model selection criterion, we found the negative binomial regression performs better than the Poisson regression. Detailed analysis indicates that the higher web-search activity of 2016 related to the sci-indexed website has a positive significant impact on the research publication of 2017. We observed that unique collaborations of 2016 and web-search activity of 2016 have a non-linear but significant positive impact on the research publication of 2017. What-if analysis indicates the high web browsing activity leads to more number of the publication. However, interestingly we see a scientist with low web activity can be as productive as others if her/his maximum hits are the sci-indexed journal. That is if the scientist uses web browsing only for research-related activity, then she/he can be equally productive even if her/his web activity is lower than fellow scientists.
\end{abstract}


\section{Introduction}
 
  %Scholarly communities are the groups of scientists who are active in scientific investigation in a research institute. 
  Research publications are one of the measures of the output of a research institute. In this paper, our objective is to find the factors which cause the scholarly communities to engage in active research and  publications. Nowadays, researchers heavily rely on digital libraries like science direct, research gate, the web of science, etc.  In this study, we try to understand how the web activity of scientists causes an impact on their research publication. The problem of discovering causal impact is ubiquitous in biological science, economics, public policy, and many regard of our daily life requiring logical reasoning and decision-making, see e.g., \cite{Judea2009, Isabelle12008}.  Thus, causal questions are about the mechanism behind the data or about predictions after a novel intervention is applied to the system, see \cite{Marloes2015}.   
   
   When it comes to learning causality from data, we should be careful about the differences between statistical associations and causations. In \cite{Ruocheng2010}, presented the example that when the weather temperatures are high, the owner of an ice cream parlor may see high electric bills along with high sales. It means there would be a strong association between the electricity bill and the revenue. However, the high electricity bill did not cause high sales. In this case, the weather temperature is the common cause of both the high electricity usage and the high sales numbers.   We say that temperature is the confounder of the causality of the electricity usage on ice cream sales. The standard statistical analysis focuses on correlation and not necessarily addresses the issues of the causal inference, see \cite{judea2010}.
 
As computing systems are integrating with the daily lives of people, it is important to understand the causal effects of these interventions correctly. The digital world generates a staggering amount of data. While these massive data sets unlocked novel opportunities to understand the scientific issues, still there is greater potential for research and practice, especially for causal inferences.  The \cite{Stephen2018} presented review which explores several key issues that have arisen around big data related to public health. Recently, a lot of progress happened in developing statistical causal inference tools (see, e.g., \cite{Judea2009, peters2017, VanderWeele2015}) which enable scientists to assess causal hypotheses and learn the corresponding effects empirically. Several causal inference methods, such as causal mediation approaches \cite{Pearl2012}, propensity score techniques (see e.g.,\cite{Harder2010, von2014}),  and sensitivity analyses \cite{Liu2013}, are already part of the regular methodological causal inference toolbox of researchers. 
 
 However, existing methods may still be too simple to answer complex questions of causation. In particular, the research questions that go beyond the main effects of an experimental study, like A-B testing \cite{cathyonei2014}. There are three issues which are often not adequately addressed by standard statistical tools.  For example, (i) the causation due to latent variables, (ii) complex network of causation, (iii) the iterative and dynamic processes of both, the temporal change in individual behavior \cite{Beltz2016} and change in factors at community level that affect the sustainability of evidence, see e.g., \cite{Butera2014,Muthén2015,von2014, Wolfgang2019, Imai2013, Beltz2016} and \cite{Chambers2013}. Therefore we should pay much more attention to designing proper study for causal inference.
 
 Conventional Statistical Machine Learning (SML) methods are insufficient for causal analysis; because these algorithms are built on pattern recognition, correlation, and focuses on prediction, see \cite{Amit2018}. For instance, in policy-making, one may want to use an SML algorithm to predict the number of papers that will be published by a scientist in the future. However, that may lead to posing a target in front of the scientist. As a result, it might create a negative feedback loop and compromise the quality of the publication. The \cite{cathyonei2016} presented several case studies which showed how SML based predictive models create negative feedback loops and dis-empower the members of the society under consideration. The need to adapt the practices and policies by the institute, using the available massive amounts of observational data, prompted the requirement for causal discovery. Hence, for a research institute, it is more useful to discover the factors that lead to research publication, rather than predicting the number of papers for scientists. For example, the causal inference might find factors like `collaboration' as a success for research output. This kind of insight will lead the academic society to take more initiative which will help the young researcher to take part in more collaborations. This kind of policy will be more empowering for the members of the scientific community, rather than setting them up for failure by giving them the target of the number of papers to publish.

The rest of the paper is structured as follows. In Section \ref{sec_related_research}, we present the related research. Section \ref{sec_Data_Set_Decribtion}, we describe the data set, data privacy policy and the study design. We present the exploratory data analysis in Section \ref{sec_EDA}. We present the research methodology in Section \ref{sec_research_method}. In Section \ref{sec_result}, we discuss the results and analysis of our findings. Finally, we conclude the paper with a few summary remarks in Section \ref{sec_discussion}.

\section{Related Research}\label{sec_related_research}

Several studies reported a positive relationship between collaboration and research publications, see, e.g., \cite{LinHea2009, Marcus2011, Fereshteh2013, Claudia2013, Giovani2017}. In \cite{Marcus2011}, looked into email communication and research productivity. To the best of our understanding, there is no literature which looks into web-search activity of scientists and their research productivity.

The experimental results in \cite{LinHea2009}, studied the effect of collaboration on the publication productivity of 65 bio-medical scientists at a New Zealand university over 14 years. The findings from the paper suggest that international collaboration has a positive effect on a scientist's research publications than domestic collaborations. Moreover, collaborations are linked to the article's quality.  Since \cite{LinHea2009},  used longitudinal design, the study establishes the causal effect of scientific collaboration on research publication for bio-medical scientist beyond a reasonable doubt. However, \cite{LinHea2009},  did not explore the effect of internet usage by the scientist on the research publication, which we highlighted in this paper.

Also, \cite{Marcus2011} found that the strong association between email communication and research productivity. \cite{Marcus2011} used email counts as a proxy for internet usage and found a positive association with research productivity. However, \cite{Marcus2011} did not address the causality through their design study. Therefore it indicates the association and not causation. For example, a scientist might be engaged in organizing conference or admin work, and that has nothing to do with research productivity. \cite{Marcus2011} uses the negative binomial regression models over the Poisson model, and they made an ad-hoc choice for the negative binomial model instead of using a statistical model selection criterion, like Akaike Information Criterion. \cite{Marcus2011}  reported that the lack of evidence between research productivity and scientific collaboration, which is counter-intuitive.  However, \cite{Marcus2011}  found the association between publication productivity, profession network size, and degree of collaboration.  Since \cite{Marcus2011} did not address the design issue of the causal study, therefore the study lacks to find evidence between scientific collaboration and publication productivity.  In this paper, we found a strong association between scientific collaboration and publication productivity.

The \cite{Fereshteh2013} reported that the impact factor of research publications is associated with factors like different dimensions of collaborations. For example,  the factors like the individual, institutional and international collaboration; journal and reference impacts; abstract readability; reference and keyword totals; paper, abstract and title lengths are associated with the impact of research publication. The findings of  \cite{Fereshteh2013}  suggested that the collaboration and journal are significantly associated with the higher citation. The result of  \cite{Fereshteh2013} suggested that researchers should include relevant references, extended abstracts, and engage in the widest possible working team.  Also, \cite{Fereshteh2013} reported that international collaboration has a high impact on research publications. In summary, \cite{Fereshteh2013} reported that different aspects of collaboration are significantly associate with research productivity like the citation.

The \cite{Claudia2013}  studied and reported how network embeddedness of scientists affects research output and impact of scientists.  The results of \cite{Claudia2013} indicate that the network dynamics of collaboration behind the generation of quality output contrasts dramatically with that of quantity.

In \cite{Giovani2017} discussed the relationship between the different types of collaboration and research productivity. In particular, \cite{Giovani2017} showed that only collaboration at the intramural and domestic level has a positive effect on research productivity and all forms of collaboration are positively affected by research productivity.

\section{Data Sets and Study Design }\label{sec_Data_Set_Decribtion}

\subsection{Data Set Description}
We implemented the study for scholarly communities of the  Indira Gandhi Centre for Atomic Research (IGCAR). IGCAR is one of the premier research institutes in India.  We divided the research activities of the community in four branches, namely Physical Sciences, Chemical Sciences, Engineering Sciences, and Other Sciences.  We allotted atmospheric, earth, and general sciences into  other sciences. There are 262 members in the community who have published at least one paper in the year 2016. We analyze the factors of those members, which leads them to publish articles in the year 2017. The factors may be their demographic data like years of experience in the institute, research branch, the number of collaborations in the year 2016, etc. and their internet activity like the amount of time spent in browsing scientific articles, etc.  
The database contains data set from four sources: (i) Publication database of 2017, (ii) Publication database of 2016, (iii) Demographic Database and (iv) Weblog data set of 2016. 

\vspace{0.3cm}

\textbf{Variable of Interest}: The number of the research publication on 2017, from the Publication database of 2017, is our target variable of interest, and we want to test which are the predictors that affect the number of the research publication on 2017.

\vspace{0.3cm}

\textbf{Demographic Predictors}: We considered demographic variables like, (i) whether the member is Ph.D. or not, (ii) the years of experience of the member, (iii) the branch of research interest. 

\vspace{0.2cm}

\textbf{Publication Database of 2016}: It consists of publications of all the 262 members. We derived the number of collaborations made by the scientist in the year 2016 from this database. For example, if a scientist A publishes two papers. The first paper with B and C; and the second paper with C and D; then we consider that A has two unique collaborations. Similarly, we counted the collaborations of the 262 members in the year 2016. In this technique, if three scientists A, B, and C publishes four papers together, then we count them as one collaboration.

\vspace{0.2cm}

\textbf{Weblog variables of 2016 }:
The access to e-journals gets dissipated in large amount in the form of logs in the web servers. We considered the variable like number of hits;   the time  spent on browsing scientific articles; the number of hours spent in the weekends; the number of visits on the top viewed journal group like IEEE, Science Direct, Springer, Elsevier, Taylor and Francis, etc. by the  members;  the number of maximum hits made in the particular month; the number of hours spent in the morning/evening hours; the  month where maximum hits made, and the maximum download size by the members.

\subsection{Data Privacy}

  We committed to the privacy of the member of the society under study. Hence we considered data privacy as an essential part of our work. We employed Data Masking to preserve data privacy of the members of the society under study. Randomly generated identifiers masked all personal information from all sources  such as names, mail ids, gender, and departmental groups disguised in all databases.

\subsection{Study Design}
\begin{figure}
        \centering
        \begin{tabular}{cc}
            \includegraphics[width=8.5cm]{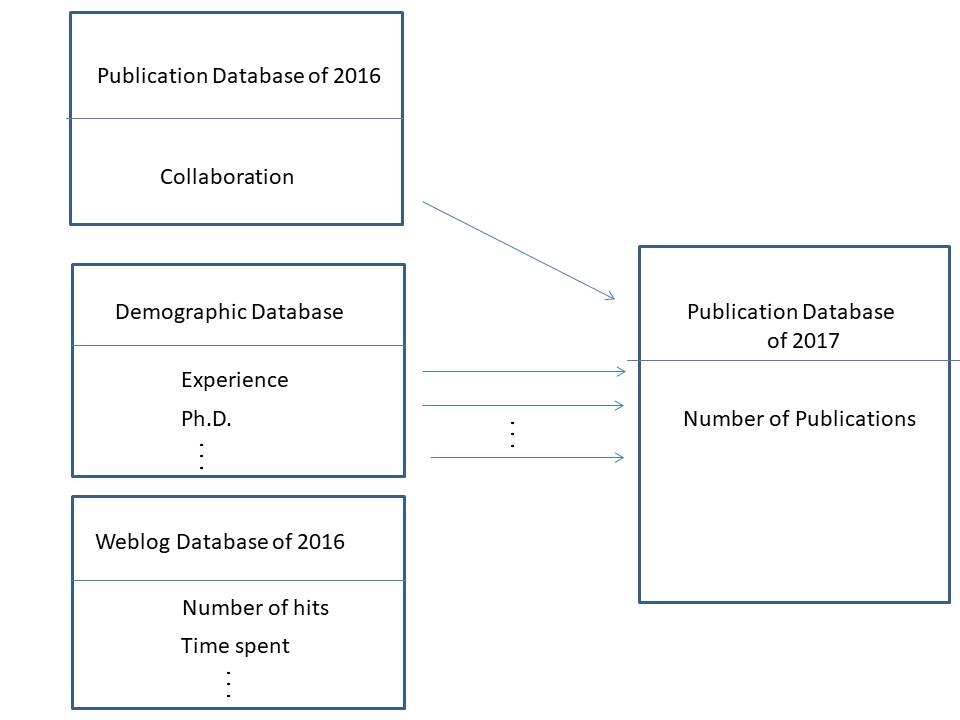} &\includegraphics[width=8.5cm]{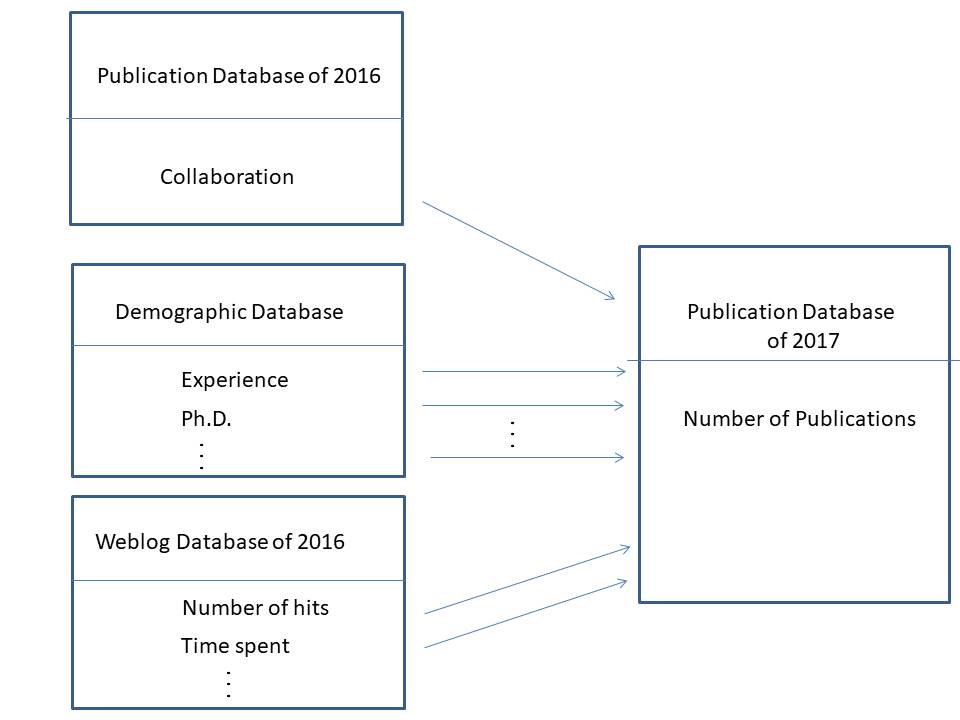}\\
            Null Study Design & Alternate Study Design
        \end{tabular}
        \caption{Study Design for Causal Impact of Web Browsing on Publication Database. The null study design shows there is no edge between weblog variables of 2016 and the publication database of 2017. It means weblog variables of 2016 does not have any impact on the publication database of 2017.  The Alternate study design indicates weblog variable of 2016 does have an impact on the publication database of 2017.}
        \label{fig_study_design}
    \end{figure}

One must be careful to establish the causal connection between the variable of interest and the estimated effect of the factors.  The co-variation is a necessary but not sufficient condition for causal inference, see e.g., \cite{tufte2006}. Correlation is not causation. However, it is a good sign that causation may exist. We often forget that the causal association must be established by design and should not rely upon statistical models whose postulates seldom defended, see e.g., \cite{neyman2007}. We presented our study design in  Figure \ref{fig_study_design}, where `Publication Database of 2016', `Weblog Database of 2016' and `Demographic Database' considered as factors; which might have a causal impact on the `Publication Database of 2017'. We assumed the `Demographic Database' of the scientist of IGCAR as constant over 2016 and 2017. However, we understand the `Demographic Database' is dynamic over a long period, which is out of the scope from the current analysis.  The `Publication Database of 2016' surely have an impact on `Publication Database of 2017'; because most of the IGCAR scientists are working in long-run collaborative projects. Our objective is to identify if the `Weblog database of 2016' has any effect on the `Publication Database of 2017'; after taking care of the impact from `Publication Database of 2016' and `Demographic Database.'

\section{Exploratory Data Analysis}\label{sec_EDA}

\begin{figure}[ht]
        \centering
        \includegraphics[height=7.5cm]{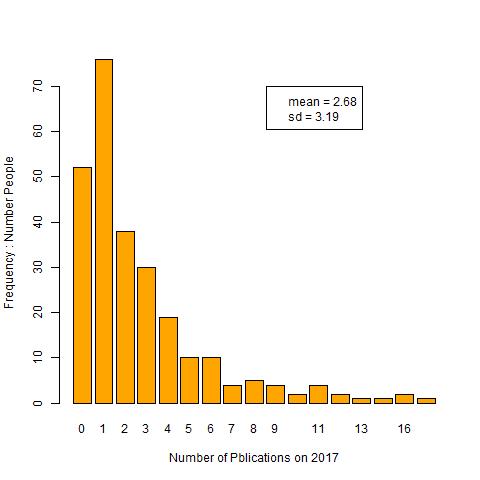}
        \caption{Bar plot for Number of publications  on 2017 vs number of scientist. The average number of publication by the scientists are 2.68 and the standard deviation is 3.19.}
        \label{fig_barplot}
    \end{figure}

In Figure \ref{fig_barplot}, we present the number of publication in 2017, versus the number of scientists who published at least one paper in 2016. Figure \ref{fig_barplot} indicates the distribution of the number of publication has a decaying effect. Hence a Poisson probability model or negative binomial model would be an appropriate probability model, for the `number of the publication.' From Figure \ref{fig_barplot},  we see that the mean (2.68) is less than the standard deviation (3.19); which indicates that the Poisson model  may not be the appropriate model. However, the negative binomial could be a more appropriate model as the variance of the negative binomial model is larger than the mean.

\begin{figure}[ht]
        \centering
        \includegraphics[height=7.5cm]{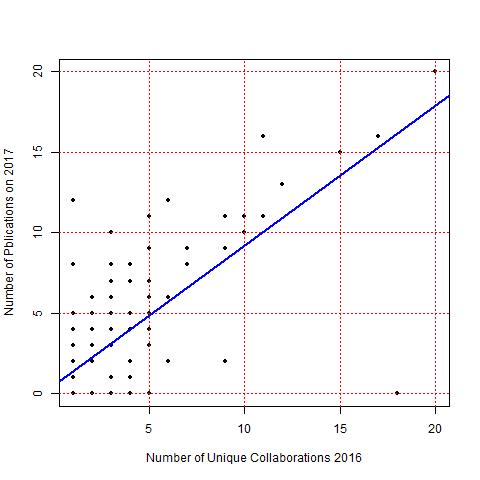}
        \caption{Plot showing the relationship between total publications on 2017 and unique collaborations in 2016. Higher the number of collaborations in the past year, higher is the number of publications in the current year.}
        \label{fig_pub_vs_collaboration}
    \end{figure}

\begin{figure}[ht]
        \centering
        \includegraphics[height=7cm]{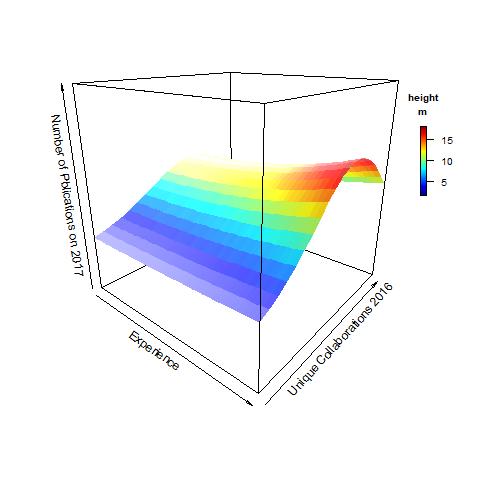}
        \caption  {3D surface plot showing the relationship between  Experience, Unique Collaborations of 2016 and Publications on 2017. A non-linear relationship exists between the three variables. }
        \label {fig_3d_surface_plot}
    \end{figure}

\begin{figure}[ht]
        \centering
        \includegraphics[height=7.5cm]{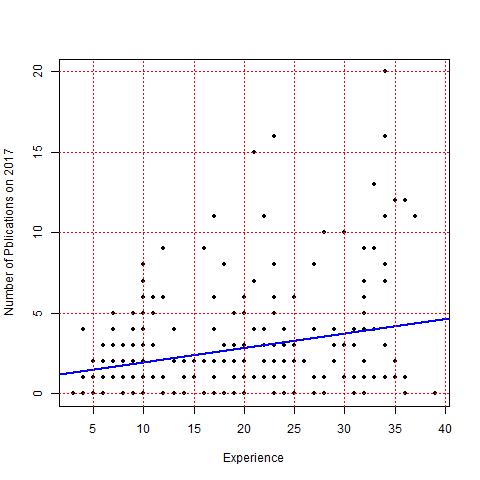}
        \caption{Plot showing the relationship between the number of publications on 2017 and years of experience of the scientists. With increase in the years of experience there is an increase in the number of publications of the scientists.}
        \label{fig_pub_vs_experience}
    \end{figure}

In Table \ref{tab_average_median_number_of_pub}, we present the average and the median number of publication by the scientist with a Ph.D. and without Ph.D. The table indicates that the scientist with Ph.D. tends to publish two more paper than their colleague without Ph.D. Figure \ref{fig_pub_vs_experience} presents the years of experience of the scientist and the number of publication in 2017. The trend and variability increase with years of experience. Figure \ref{fig_pub_vs_collaboration} presents a strong positive relationship between the number of a unique collaboration in 2016 and the total number of publications in 2017. This strong positive association is along the same line of the available literature of last decade, see \cite{LinHea2009, Marcus2011, Fereshteh2013, Claudia2013, Giovani2017}. Finally, in Figure \ref{fig_3d_surface_plot}, we plot the 3D surface between years of experience, unique collaboration of 2016 and the number of publication of 2017. The plot indicates the existence of possible non-linear behavior between the three variables.

 %\begin{figure}[ht]
 %       \centering
 %       \includegraphics[height=7.5cm]{boxplot_for_toal_pub_2017_vs_PhD}
 %       \caption{No. of publications vs Ph.D.}
 %       \label{fig_boxplot_pbulication_vs_phd}
 %   \end{figure}

    %\begin{figure}[ht]
    %    \centering
    %    \includegraphics[height=7cm]{3d_scatter_plot}
    %    \caption  {3D scatter plot of Experience Vs Unique Collaborations Vs Publications}
    %    \label {fig_3d_scatter_plot}
    %\end{figure}

\section{Research Methodology}\label{sec_research_method}

In this section, we present the two models and the rationale for formulating the null and alternative hypothesis for our analysis under the proposed models. We also present the underlying assumptions of the models and analysis.

\subsection{Assumptions Related to Causal Inference}

We follow \cite{imbensbook} and made the following assumptions throughout our study.

\begin{enumerate}
    \item \textbf{Unconfoundedness}: The Assignment is free from dependence on the potential outcome. That is the web activity of a scientist in 2016 will not be dependent on how many publications she/he is going to have in 2017.
    \item \textbf{Individualistic Assignment}: The assignment mechanism is individualistic. The probability of sample unit $i$ (i.e., a randomly chosen scientist) is a function of pre-treatment variables for the unit $i$ only and free of dependence on the values of pre-treatment variables for other units. That is a web activity of a randomly chosen scientist is independent of the web activity of any other scientist of the community.
    \item \textbf{Probabilistic Assignment}: The assignment mechanism is probabilistic so that the probability of receiving any level of the treatment is strictly between zero and one. In other words, the probability of the web activity of a scientist is purely random in nature.
\end{enumerate}

\subsection{Probability Models}

\subsubsection*{Poisson Regression Model}

In this study, we consider the measure of research output ( i.e., the total number of publications 2017) as a count variable. The most popular probability model for count variable is Poisson distribution and probability of the number of publications can be modeled as
$$
P(Y=k)=e^{-\lambda}\frac{\lambda^k}{k!}~~~k=0,1,2,\cdots,
$$
where $Y$ denote the number of publications in 2017 and $\lambda$ is the rate of publication in 2017. The rate of publication can be modeled as
$$
\log(\lambda) = \eta = \beta_0+\beta_1X_1 + \beta_2X_2 + \cdots + \beta_pX_p,
$$
where $X_1,X_2,\cdots,X_p$ are the predictor variables, such as the number of a unique collaboration in 2016, scientist's years of experience, if the scientist has a Ph.D. (or Not),  web-search activities, etc. 

\subsubsection*{Negative Binomial Regression Model}

We can use the negative Binomial probability model as an alternative model for the total number of publications 2017 in the following way,
$$
P(Y=k)={{r+k-1}\choose{k}}p^r(1-p)^k,~~k=0,1,2,3,\cdots
$$
where the predictor variables can be modeled as
$$
\log\Big(\frac{r(1-p)}{p}\Big)= \eta = \beta_0+\beta_1X_1 + \beta_2X_2 + \cdots + \beta_pX_p.
$$
The regression coefficients of the Poisson and Negative Binomial model can be estimated using standard tools, see e.g., \cite{DasDey2006,DasDey2013}.

\subsection{Hypothesis}

\subsubsection*{Null Hypothesis $H_0$} 

The last decades of research, see e.g., \cite{LinHea2009, Marcus2011, Fereshteh2013, Claudia2013, Giovani2017}, establishes beyond a reasonable doubt that collaboration is the main factor of research productivity. Hence, we assume the number of publications on 2017 (the measure of research productivity), is only the function of the number of collaborations in 2016, and other demographic variables like years of experience, and  Ph.D.'s or not. Our exploratory data analysis also indicates the same. Under the null model, the web-log variables like scientist web-search activity of 2016 have no impact on research productivity of 2017. Hence we have the null model as
\begin{eqnarray*}
\eta= \beta_0 &+&\beta_1 \text{  Ph.D.'s or not } + \beta_2  \text{ experience } \\
    &+&  \beta_3  \text{ unique collaborations in 2016 } \\
    &+&  \beta_4 \text{ experience *  unique collaborations in 2016} 
\end{eqnarray*}

\subsubsection*{Alternative Hypothesis $H_A$}

We formulate our alternative hypothesis as follows. In addition to the number of collaborations in 2016 and other demographic variables, the web-log variables like scientist's web-search activity of 2016 do have an impact on research productivity of 2017. For example: If the number of maximum hits by a scientist in sci-indexed journals, has an effect on the research output of 2017. Hence we have the alternative model as
\begin{eqnarray}\label{eqn_alt_model_NB}
\eta= \beta_0 &+&\beta_1 \text{  Ph.D.'s or not }  +  \beta_2  \text{ experience } \nonumber \\
    &+&  \beta_3  \text{ unique collaborations in 2016 } \nonumber \\
    &+&  \beta_4 \text{ experience *  unique collaborations in 2016} \nonumber  \\
    &+&  \beta_5 \text{ hits new} + \beta_6 \text{ sci-indexed Journal} \nonumber \\
    &+& \beta_7 \text{ download doc size} +  \beta_8 \text{ download doc size}^2\nonumber \\
    &+& \beta_9 \text{ unique collaborations in 2016 * hits new} \nonumber \\
    &+& \beta_{10} \text{unique collaboration 2016 * (download doc size)}^2
\end{eqnarray}

\section{Results and Analysis}\label{sec_result}

In Table \ref{Tabl_AIC_Pois_NB}, we present the Akaike's Information Criterion (AIC) for both Poisson and Negative Binomial Regression under Null and Alternative model. AIC for Negative Binomial Regression is smaller than Poisson regression, indicates that the Negative Binomial Regression performs better than Poisson regression. Hence now onwards, we present all our analysis only based on the Negative Binomial Regression. Note that \cite{Marcus2011} reported the negative binomial regression is a better model compare to Poisson regression. Hence our findings are in-line with previous findings of \cite{Marcus2011}. The findings from \cite{Marcus2011} were based on agricultural scientist of Philippine. However, our findings are based on the research productivity of Indian scientist. The two independent studies indicates that the negative binomial regression perhaps a good model for studying research productivity.

In Table \ref{Tabl_LR_test_alt_modl}, we present the Likelihood Ratio (LR) based Chi-Square Test between Null and Alternate Model. The P-value for the test indicates that we reject the null model in favor of the alternate model at 0.01\% level of significance. It indicates that the web-search activity of 2016 has a statistically significant effect on the number of  research publication for 2017. Note that the LR test in the Table \ref{Tabl_LR_test_alt_modl} only indicates that web-search activity of 2016 has a significant effect over research publication of 2017. But it does not say anything about the direction. Hence we present further analysis.

 Based on analysis of alternate negative binomial regression model in Equation \ref{eqn_alt_model_NB}, presented in Table \ref{tabl_analysis_alt_model_NB}; the demographic variables like Ph.D., Experience and unique Collaborations of 2016 have a statistically significant impact on the number of research publications of 2017. This supports the findings of \cite{Fereshteh2013}. In addition to demographic variables, the log variables of 2016 like viewing scientific indexed journals, document size of the downloaded documents, etc. are also statistically significantly affect the research publications of 2017. The positive value of the coefficient estimate for log-variables of 2016 like `hits new', `sci indexed journal sites', `download document size' indicate,  these variable have a positive significant impact on research publication of 2017. Note that the coefficient of interactions between the variables is significant, however, the negative value of interaction indicates the system is complex. Hence in order to understand the effect of the proposed complex system, we present three scenarios in `What-if' analysis.

\subsection*{What-if Analysis}

We considered three scenarios. In Table \ref{Tabl_Summary_Stat_Web_variables}, we present the first, second and third quartile of web browsing activity. Where we considered the total number of hits and size of the downloaded document as representative features for web browsing activity. If a scientist's number of hits and size of downloaded document are both at the first quartile level, then we consider that scientist as one who is less active in web browsing. Similarly, if a scientist's both numbers of hits and download document size are at the third quartile level, then we consider the scientist as one with a high level of web-activity. 

Sci-Index = 1 indicates that the scientist's maximum hit is the website of some sci-indexed journal and Sci-Index = 0 indicates that the scientist's maximum hit is the website which is not a site of a Sci-Index journal. Median for Sci-Index is 1, means more than 50\% scientist's maximum web-search activity is Sci-indexed journal. 

We considered the demographics of two scientists are exactly same. We assumed two scientist having Ph.D., five collaboration on 2016 and 10 years of experience.

In all three scenarios the first scientist were always having high web activity and maximum hit of the sci-indexed journals. We presented the result of the what-if analysis in Table \ref{tabl_what_if_analysis}. We estimated the standard error and $95\%$ confidence interval, using bootstrap statistics with 5000 bootstrap samples.

\subsubsection*{Scenario I}

In the first scenario, we considered the second scientist has the low web activity and her/his maximum hit is not the site of the sci-indexed journal. The difference in the expected number of publication between the two scientists is 1.62 with a standard error of $0.6$ and $95\%$ confidence interval $(0.47,2.83)$. This indicates that the difference between the two scientists is statistically significant and one with high web-activity publishes at-least one more paper than the scientist with low levels of web-activity.

\subsubsection*{Scenario II}

In the second scenario, we considered the second scientist has a high web activity and her/his maximum hit is not the site of the sci-indexed journal. The difference in the expected number of publication between the two scientists is 1.35 with a standard error of $0.56$ and $95\%$ confidence interval $(0.22,2.45)$. This indicates that the difference between the two scientists is statistically significant. Now as both scientists are at the high level of web-activity. But the second scientist's maximum number of hit is not the sci-indexed journal. It means the main feature that differentiates between the two scientists is if the maximum number of hit is the sci-indexed journal.

\subsubsection*{Scenario III}

The second scenario leads us to check the third scenario. In the third scenario, we considered the web-activity of the two scientists, same as that of scenario I. That is the first scientist has the high level of web-activity and the second scientist have the low activity. However, both scientists number of maximum hit is the sci-indexed journal. Now the difference reduces to 0.35 with $95\%$ confidence interval $(-0.60,1.37)$ includes zero. It indicates that the second scientist who has a low level of web-activity is as effective as the scientist with a high level of web-activity because the scientist's maximum number of hits is sci-indexed journal site. It means a scientist can have less web-activity but if the web-activity is only related to research then that scientists productivity will be as high as another.

\section{Discussion}\label{sec_discussion}

%\subsubsection*{Data Privacy and Challenges}

% The private information about all the members of the society under consideration was deleted and id was masked under randomly generated identifiers. 

%When we try to find the impact of download document size behavior on the research outputs, we found an outlier with extreme download size behavior in Figure \ref{figure_outlier_download}. For further analysis, we decided not to infringe the privacy of the member of our society and not studied the impact of download size. 

The concept of causation has long been controversial in qualitative research, and many qualitative researchers have rejected causal explanation as incompatible with an interpretive or constructivist approach \cite{Joseph2012}. Unable to analyze the quality of the publications. In the future, we propose to measure the quality of the publications through citations. Causal inferences eliminate the possibility of the bad feedback loop involved in the case of predictive modeling models. Instead, it discovers useful inferences which can be adapted as policies in the institute for more publications. It helps to assist the members of our society and not directing them with rules which are difficult to comply.  Also, we tried to preserve the privacy of the members of our society.

Our analysis indicates that the Negative Binomial Regression performs better than Poisson regression. Perhaps it captures overdispersion exists in the data. The demographic variables like Ph.D., Experience and unique Collaborations of 2016 have a statistically significant impact on the number of research publications of 2017. In addition to demographic variables, the web-log variables of 2016 like viewing scientific indexed journals,  document size of the downloaded documents,  etc.  shows statistically significant effect on the research publications of 2017. 

What-if analysis indicates the web browsing activity leads to more number of the publication. However, interestingly we see a scientist with low web activity can be as productive as others if her/his maximum hits are the sci-indexed journal. That is if the scientist uses web browsing only for research-related activity, then she/he can be equally productive even if her/his web activity is lower than fellow scientists.

\bibliographystyle{abbrvnat}
\bibliography{bibliography_first_paper}

\newpage

 %\begin{figure}[ht]
 %       \centering
 %       \includegraphics[height=7.5cm]{plot_challenge_outlier_data}
 %       \caption{Download Doc Sixe vs. Number of Hits. The plot shows an outlier in the data}
 %       \label{figure_outlier_download}
 %   \end{figure}

\begin{table}[ht]
        \centering
        \begin{tabular}{c|c|c|c}\hline
           Status  & Sample size & Mean & Median \\ \hline
           PhD  & 103 & 3.91 & 3 \\
           No PhD & 158 & 1.88 & 1\\ \hline
           Total & 261 & 2.68 &2 \\ \hline
        \end{tabular}
        \caption{Average \& Median values of the number of publications by Ph.D.'s and non Ph.D.'s }
        \label{tab_average_median_number_of_pub}
    \end{table}

\begin{table}[ht]
\centering
\begin{tabular}{l|r|r}
  \hline
 & Null Model ($H_0$) & Alternate Model ($H_A$) \\ 
  \hline
Poisson Regression & 1040.96 & 993.77 \\ 
  Negative Binomial Regression & 994.52 & 974.53\\ 
   \hline
\end{tabular}
\caption{ Akaike's Information Criterion (AIC) for both Poisson and Negative Binomial Regression under Null and Alternative model. AIC for Negative Binomial Regression is smaller than Poisson regression indicates that Negative Binomial Regression performs better than Poisson regression. }
\label{Tabl_AIC_Pois_NB}
\end{table}

\begin{table}[ht]
\centering
\begin{tabular}{l|c|c|c|c}
  \hline
 &  Residual degrees of freedom &  2 x log-like & LR stat. & P-value \\ 
  \hline
Null Model & 256 & -982.52 &  &  \\ 
Alternate Model  & 250 & -950.53 & 31.99 & $<$0.0001 \\ 
   \hline
\end{tabular}
\caption{Likelihood Ratio based Chi-Square Test between Null and Alternate Model under Negative Binomial Regression. The P-value for the test indicates that we  reject null model in favour of the alternate model at 0.01\% level of significance. It indicates that web-search activity of 2016 has statistically significant effect on the number of research publication for 2017.}
\label{Tabl_LR_test_alt_modl}
\end{table}

\begin{table}[ht]
\centering
\begin{tabular}{lcccl}
  \hline
 & Estimate & Std. Error & z value & P-value \\ 
  \hline
Intercept & -7.968e-01 & 1.967e-01 & -4.050 & 5.11e-05 \\
PhD       & 2.315e-01  & 1.073e-01 &  2.158 &  0.0310 \\
experience & 3.126e-02 & 7.468e-03 &  4.186 & 2.84e-05 \\
unique  collaboration  2016 & 4.282e-01 & 5.831e-02 &  7.343 & 2.08e-13\\
hits new  & 1.885e-05 & 4.330e-06  & 4.353 & 1.35e-05 \\
sci indexed & 2.520e-01 & 1.085e-01 & 2.323 & 0.0202 \\
download doc size & 2.414e-08 & 1.587e-08 & 1.521 &  0.1283 \\ 
$I$(download doc size$^2$)  & -2.248e-15 & 1.005e-15 & -2.238  & 0.0253 \\  
experience$:$unique collaboration 2016 & -8.146e-03 & 1.887e-03 & -4.318 & 1.58e-05 \\
 
unique collaboration 2016$:$hits new & -5.521e-06 & 1.224e-06 & -4.511 & 6.45e-06 \\
unique collaboration 2016$:$$I$(download doc size$^2$) & 5.756e-16 & 2.657e-16 &  2.167 &  0.0303 \\ \hline
\end{tabular} 
\caption{Coefficient Estimates, standard error, z-vale and P-value for the Negative Binomial Regression mode presented in the equation (\ref{eqn_alt_model_NB})}
\label{tabl_analysis_alt_model_NB}
\end{table}

 %\begin{verbatim}
  %          weekend     hits sci_indexed download_doc_size
%Min.              0        0           0                 0
%1st Qu.           0     2268           0           1084444
%Median           30     4903           1           2872888
%3rd Qu.         306    10163           1           5632543
%Max.           4535   985370           1         128580955

%\caption{table showing 1st and 3rd quartile values of the internet activity of scientists}
 %        \end{verbatim}

    \begin{table}[ht]
        \centering
        \begin{tabular}{l|rr|r}\hline
             & \multicolumn{2}{c|}{Web Activity} & Max hit \\
             &  Hits  & Download Doc Size & Sci-Indexed\\ \hline
    First Quartile &  2268 & 1084444  & 0 (= No ) \\
    Median  &   4903 &  2872888  & 1 (= Yes) \\
    Third Quartile & 10163 & 5632543 & 1 (= Yes)\\ \hline
        \end{tabular}
        \caption{Summary Statistics of three variables of web-search activity. Sci-Index = 1 indicates that the scientist's max hit is the website of some sci-indexed journal and Sci-Index = 0 indicates that the scientist's max hit is the website which is not a site of a Sci-Index journal. Median for Sci-Index is 1, means more than 50\% scientists maximum web-search activity is Sci-indexed journal.}
        \label{Tabl_Summary_Stat_Web_variables}
    \end{table}

    \begin{table}[ht]
        \centering
        \begin{tabular}{llllccc}\hline
            & Scientist & Web & Max hit & Expected Number  & Standard  & 95\% CI \\
            & &  activity& Sci-Index & of Publication & Error & \\ \hline
           Scenario I & 1&High  &  Yes & 6.05 & 0.96 & (4.35,8.15)\\
             & 2& Low& No& 4.43 & 0.93 & (2.74,6.41)\\ 
         & Difference &-&-&1.62& 0.60& (0.47,2.83)\\     
         \hline\hline
          Scenario II  & 1&High & Yes & 6.05 & 0.96 & (4.35,8.15) \\
             & 2&High & No & 4.70 & 0.91 & (3.09,7.70)\\ 
             &Difference&-&-&1.35& 0.56 & (0.22,2.45)\\     \hline \hline
          Scenario III  & 1&High &Yes & 6.05  & 0.96 & (4.35,8.15)\\
             & 2 & Low& Yes & 5.70 & 1.07 & (3.76,7.98)\\ 
        &Difference&-&-&0.35& 0.49 & (-0.60,1.37)\\     \hline \hline 
          \end{tabular}
        \caption{What-If Analysis of 3 Scenarios. In each scenario the first scientist's web activity is high and the maximum hit made by the scientist  is the site of the sci-index-ed journal. For the second scientist we compare different combination of web-activity and Max-hit on Sci-Index journal. In Scenario I and II, the difference is more than one publication and the $95\%$ CI indicate the difference is statistically significant as it does not include 0. However, for the scenario 3, the difference is not significant. It means one can have low web-activity, but if the maximum hit is sci-indexed journal, then that scientist will be as equally effective in research publication. We estimated the standard error and $95\%$ confidence interval, using bootstrap statistics with 5000 bootstrap samples.}
        \label{tabl_what_if_analysis}
    \end{table}
    
     %\begin{table}[ht]
     %        \centering
     %        \begin{tabular}{c|cc}\hline
     %                          & Null Model  & Alternate\\ \hline
     %        Scientist with     &             & \\
     %        low web activity  &     5.61         &  4.43 \\ \hline
     %        Scientist with     &             & \\
     %        high web activity  &     5.61         & 6.05\\ \hline
     %        \end{tabular}
     %        \caption{ Results of Scenario Analysis I. Expected Number of Publication  of the scientists with high and low internet activity}
     %        \label{tab_scenario_1}
     %    \end{table}

      %     \begin{table}[]
      %       \centering
      %       \begin{tabular}{c|cc}\hline
      %                         & Null Model  & Alternate\\ \hline
      %       max hit with    &             & \\
      %       non sci-index  &     5.61         & 4.70 \\ \hline
      %        max hit with     &             & \\
      %        sci-index   &     5.61         & 6.05 \\ \hline
      %       \end{tabular}
      %       \caption{ Results of Scenario Analysis II. Expected Number of Publication  of the scientists whose max hits are with sci-index journal site. }
      %       \label{tab_scenario_2}
      %   \end{table}
         
\end{document}